# Quantifying the Performance of Federated Transfer Learning

**Qinghe Jing**[1], **Weiyan Wang**[1], **Junxue Zhang**[1,2], **Han Tian**[1], **Kai Chen**[1,3]

[1]Hong Kong University of Science and Technology
[2]Clustar Technology Co., Ltd.
[3]Peng Cheng Lab
{qjing, wwangbc, jzhangcs, htianab, kaichen}@cse.ust.hk

## Abstract

The scarcity of data and isolated data islands encourage different organizations to share data with each other to train machine learning models. However, there are increasing concerns on the problems of data privacy and security, which urges people to seek a solution like Federated Transfer Learning (FTL) to share training data without violating data privacy. FTL leverages transfer learning techniques to utilize data from different sources for training, while achieving data privacy protection without significant accuracy loss. However, the benefits come with a cost of extra computation and communication consumption, resulting in efficiency problems. In order to efficiently deploy and scale up FTL solutions in practice, we need a deep understanding on how the infrastructure affects the efficiency of FTL. Our paper tries to answer this question by quantitatively measuring a real-world FTL implementation FATE on Google Cloud. According to the results of carefully designed experiments, we verified that the following bottlenecks can be further optimized: 1) Inter-process communication is the major bottleneck; 2) Data encryption adds considerable computation overhead; 3) The Internet networking condition affects the performance a lot when the model is large.

## 1 Introduction

Modern machine learning applications cannot succeed without the availability of training data [Sun *et al.*, 2017]. The successes of applying AI technology in computer vision and e-commerce recommendation system all rely on massive well-labeled datasets, such as ImageNet [Deng *et al.*, 2009]. However, in some industries, such as healthcare [William, 1995] and finance [Yeh, 2016], training data is not easy to obtain for the following two reasons: 1) labeling raw data for machine learning training is expensive because they require professional knowledge and experience, 2) the increasingly strong protection towards data privacy and data security restricts the sharing of training data. Consequently, for those organizations or parties, they own insufficient data (few samples and features) for training a high-quality machine learning model. Furthermore, they cannot even share data with each other due to some regulations such as the General Data Protection Regulation (GDPR) [Regulation, 2016]. The regulation makes it unacceptable to collect data from various sources and integrate them into one location without user permission. This motivates people to develop frameworks which allow data sharing for machine learning training without violating data privacy.

Federated Learning (FL) is a solution to overcome the constraints. A Federated Learning (FL) system for mobile devices was first proposed by Google [McMahan *et al.*, 2016] that enables users to form a federation and train a centralized model while their data are stored securely local. However, it requires the training data owned by different parties to share the same feature space, which restricts its practicality. WeBank takes a further step by proposing their Federated Transfer Learning (FTL) [Liu *et al.*, 2018; Cheng *et al.*, 2019]. It provides a more reliable method for specific industries by applying homomorphic encryption and polynomial approximation instead of differential privacy. What's more, participants in FTL can have their specific feature space which makes FTL suitable for more scenarios.

Nowadays, FTL is attracting increasing attention in industries such as finance, medicine and healthcare. To make FTL a practical solution in real-world environment, performance is a key factor. However, FTL has its own unique privacy protection components such as data encryption, leading to potential performance degradation. The workflow of FTL is similar to distributed machine learning, while data are stored in different data owners. The ideal performance that FTL can achieve should be close to that when the data are simply owned by one party and trained in a distributed way. However, due to those data privacy requirements, the performance gap always exists. To the best of our knowledge, none has surveyed the performance of FTL, resulting in no guidance on how to deploy and optimize FTL in real environment. A comprehensive understanding will provide us with clear directions to improve the efficiency of FTL framework in terms of model training and data transmission.

In this paper, we are going to quantify the performance gap based on the real-world project FATE [WeBankFinTech, 2019]. We first measure the end-to-end task completion time to show the overall performance gap between FTL and distributed machine learning paradigm. Then we break down the task completion time into CPU execution time and data transfer time to give a detailed view of which factors cause the considerable gap. Further, we modify the hardware configurations such as virtual CPU cores and the amount of RAM to explore whether advanced configurations can mitigate the overhead in CPU execution time. We also emulate the real working scenarios of FTL where parties locate in different data centers to explore the impact of networking environment towards overall performance. Finally, we show three major bottlenecks in FTL and their potential solutions:

- Inter-process communication is the major bottleneck of current FTL implementation. Within one machine, data exchange and memory copies among processes cause extremely high latency. Techniques such as JVM native memory heap and UNIX domain sockets grant us the chance to mitigate the bottleneck.

- The requirements for privacy protection adds more computation overhead to FTL. The software-based encryption implementation consumes too much CPU cycles. Works on SmartNICs [Ovtcharov *et al.*, 2015; Firestone *et al.*, 2018] inspires us to consider the possibility of implementing data encryption on specific hardware in the FTL infrastructure.

- Traditional congestion control problems in network are also potential challenges. With the bandwidth becomes narrow, the intensive data exchange through network will make FTL frameworks face more heavy network traffic. Deploying advanced networking technologies such as PCC [Dong *et al.*, 2018] will help to make data transfer more robust.

In the following sections, we first show the background of FTL and some specific concepts in FATE. Then we present our experiment methodology including settings and the metrics we focus on. Finally, we evaluate the performance of FTL in details and propose some possible solutions towards those potential bottlenecks.

## 2 Background and Motivation

### 2.1 Federated Transfer Learning

With the growing need of data for successful machine learning applications and the increasing concerns on data privacy protection, different organizations are motivated to explore frameworks that allow data sharing without violating data privacy. Google first introduced Federated Learning (FL) system for mobile devices where individuals can form a federation and build a globally high-quality model. The private data is kept locally and encrypted models are exchanged among parties. Differential privacy is used in their privacy-preserving collaborative machine learning system [McMahan *et al.*, 2016] because it provides a probabilistic view of privacy. It adds noise on intermediate results to prevent the attacker from accessing and inferring the private information. However, it does not fully satisfy the requirement of FL due to the following reasons: 1) The added noise causes the decrease of accuracy leading to low-quality models [Aono *et al.*, 2018]. 2) It is still transferring sensitive information that provides the possibility of malicious attacks [Yang *et al.*, 2019].

Furthermore, their framework requires all participants to have data in the same feature space, which makes it impractical in the real scenarios where different participants may not share exactly the same feature space.

To overcome the above two challenges, WeBank proposed Federated Transfer Learning (FTL) framework to provide a more secure method for specific industries [Yang *et al.*, 2019]. FTL utilizes the transfer learning scheme under a privacy-preserving setting to overcome the lack of data or label, and improves the security level of data. It can be applied to the scenarios where two datasets differ not only in samples but also in feature spaces. Building neural networks similar to weakly-shared Deep Transfer Network proposed in [Shu *et al.*, 2015], FTL transfers features from different feature space to the same latent representation, which can then be used for training with labels collected by different participants.

Moreover, to avoid exposing data of participants during the back propagation, FTL adopts homomorphic encryption and polynomial approximation to ensure the privacy protection under an honest-but-curious setting, where a semi-honest adversary can only corrupt at most one of the two parties. It overcomes the defects by keeping raw data and models locally. Both of the parties encrypt their gradients before data transfer and use some techniques such as adding random masks which prevent parties from guessing the information of each other during any stages of the task. Because additively homomorphic encryption does not support operations except addition, FTL uses second-order Taylor approximation for gradient and loss computation to decompose the formula into an addition of several components. Models with polynomial approximation can still achieve a comparable accuracy compared to the ones using the original approach and the loss can converge at a competitive rate.

### 2.2 FTL Framework

In this section, we introduce concepts and the workflow of FTL based on the latest practical FTL implementation, FATE by WeBank [WeBankFinTech, 2019].

- **Three different roles of parties:** Members that join the federation are called parties, and there are three different parties in FTL: Guest, Host and Arbiter. Guest and Host are the data holders. And Guest is the one to launch the task-specific and multi-party model training, with datasets provided by both itself and Host. They are mainly responsible for computation and data encryption. Arbiter is the party that sends public-keys to Guest and Host before the tasks, which allow the data exchange between them. During the model training, Arbiter takes the responsibility to aggregate gradients and check whether the loss converges.

- **FTL workflow:** As shown in Figure 1, Guest and Host first calculate and encrypt their intermediate results locally using their own data, which are used for gradient and loss calculations. Then they send the encrypted values to Arbiter. Finally, Guest and Host acquire the decrypted gradients and loss from Arbiter to update their models. The FTL framework iteratively repeats the above steps until the loss function converges.

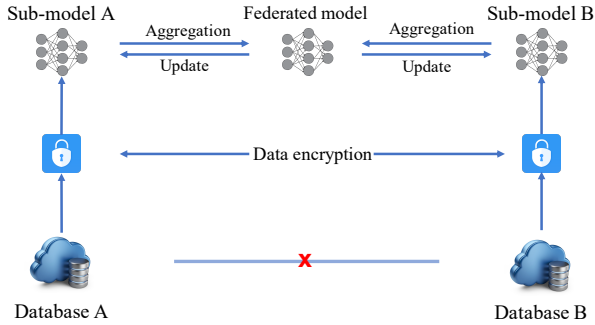

Figure 1: FTL Workflow. When training a model, party A and B first calculate and encrypt their intermedia results, including gradients and loss. Then a third party collects and decrypts the gradients and loss. Finally, party A and B receive the aggregated results and update their own model.

- **Two different training approaches:** FTL provides two different approaches for parties: homogeneous and heterogeneous. For the homogeneous approaches, parties train the same model with different samples. While for heterogeneous ones, parties share the same samples in different feature spaces, they aggregate these features in an encrypted state and build a model with all the data cooperatively.

In FATE, machine learning algorithms are implemented in Python, while the bottom framework responsible for task scheduling, thread/process management and data communication are written in Java. Within each party, there are several processors on different nodes training the model parallelly. The data transmission between different process, nodes and parties are based on the gRPC Remote Procedure Calls (gRPC) [Google, 2015].

## 2.3 Optimal Performance for FTL

FTL and distributed machine learning have some similar properties, both of them have several working nodes which holds different data and update their models according to aggregated results. However, there are also some significant differences between them. For distributed machine learning tasks, parameter server [Li *et al.*, 2014] acts as a central scheduling node, allocating data and computing resources to the working nodes to improve the training efficiency. But for FTL, data holders own the worker nodes and have full autonomy for their own data individually. What is more, FTL guarantees the data privacy of the data owners when training models, the encryption methods require more powerful computation capacity and faster network to transfer the data with increasingly large size. The requirement of data privacy protection also makes data transfer more frequent, which increases the demand for an efficient network environment. Thus FTL has a more complex framework and distributed machine learning is an ideal case for FTL where data is owned by one party and encryption is not required.

The complexity of FTL's workflow makes it a new machine learning paradigm. Without a thorough understanding of the performance of FTL, people cannot efficiently deploy and correctly optimize FTL in real production environment. However, to the best of our knowledge, little work is done to measure the performance of FTL or provide a standard for evaluation.

Therefore, we try to quantify the performance gap between FTL and distributed machine learning. Further, we break down the considerable gap into details to explore the factors that cause the performance degradation. Finally, we provide the bottlenecks and their potential solutions, which can assist the future optimization on the framework.

## 3 Methodology and Evaluation

In this section, we first present our evaluation methodology and environment. Then we show the overall performance gap between FTL and current distributed machine learning paradigm. Finally, we give a detailed view of what factors cause the considerable gap.

### 3.1 Overview

As discussed above, for a certain model and dataset, the performance of distributed training is the optimality that federated transfer training can reach. The reason is that both paradigms have similar workflow but FTL has extra overhead, such as data encryption and model modification, to guarantee the data privacy. In our paper, we try to quantify the overhead. We first measure the end-to-end task completion time, which gives us an overview of the performance gap between the two paradigms. Our results show that the gap can be over $18 \times$ (§3.2).

In order to explain what factors cause the dramatic performance gap, we further break down the task completion time into CPU time (§3.3, including computing time and copy time) and data transfer time through network (§3.4) because in both FTL and distributed machine learning, computation and data transfer play critical roles towards the performance. We now introduce the methods used to obtain the CPU time and data transfer time with high accuracy and the brief results.

For FTL, CPU time can be obtained by locating all the parties in one machine where data transfer time among them can be ignored. We use a similar method by assigning parameter server and worker in one machine to measure the CPU time in distributed machine learning. CPU execution in FTL takes an unexpected longer duration compared to distributed machine learning (225s vs 9.2s). Therefore, we take one step further by measuring how much time is spent on actual model training and how much is spent on other computation tasks such as inter-process communication and memory copy. We also conduct experiments to measure whether the data encryption in FTL also introduces large overhead. The results show that the encryption method takes $2\times$ longer time than actual model training (29s to 9s). Finally, we explore whether an advanced hardware configuration can mitigate the above overheads but the results show advanced hardware has little help on this problem CPU time reduction is within 16 % when we have different hardware configurations. We choose to take the measurement from the Guest perspective, because it is the one who initializes the request and provides the final model.

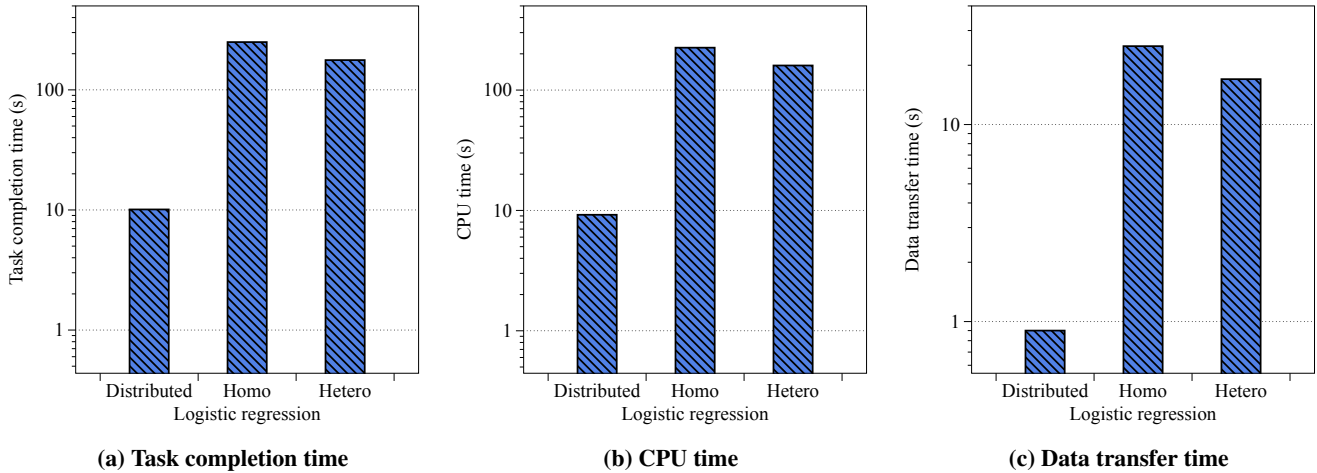

**(a) Task completion time**  **(b) CPU time**  **(c) Data transfer time**

Figure 2: Task completion time, CPU execution time and data transfer time for both homogeneous(Homo) logistic regression and heterogeneous(Hetero) logistic regression. Results show that the overall performance of FTL can be over $17\times$ longer than distributed machine learning. Both CPU execution time and data transfer time show similar trends.

Data transfer time is also an important factor because different parties in FTL exchange data more intensively and the data size is larger. We calculate the time used for data transfer through network based on the task completion time and CPU time. Data transfer in FTL costs over $18\times$ time compared to distributed machine learning (17s vs 0.9s). The result indicates that FTL has higher demands for the network environment. Furthermore, unlike distributed machine learning, parties in FTL may locate in different data centers where the networking bandwidth may not always be adequate. Thus, we use instances in different geographical levels (within one data center, within one country, within one continent and worldwide) to emulate the real working scenarios for FTL and explore the impact of networking environment towards overall performance. When the distance between parties becomes longer, data transfer time can occupy up to $34\%$ of the overall task completion time.

In our paper, we use virtual instances from Google Cloud. The dataset we use is the breast cancer dataset[William, 1995] which contains 569 samples and 32 features. To make our evaluation results representative, we perform our evaluation based on open source FTL implantation: FATE v0.1[WeBankFinTech, 2019]. We use two common logistic regression models in FTL, the homogeneous and heterogeneous LR respectively. Based on the different targets of homogeneous and heterogeneous models, the dataset is divided into two segments vertically and horizontally. The horizontal division generates different data samples which are required by the homogeneous model, while the vertical one has different feature spaces that will be used in heterogeneous model. As a comparison, we implement distributed logistic regression using TensorFlow v2.0.0 [Abadi *et al.*, 2016] with one parameter server and two workers. In all experiments, the models are trained with the same number of iterations.

### 3.2 Overall Performance

In this section, we show the overall performance comparison between FTL and distributed machine learning.

First, we measure the end-to-end task completion time of FATE by training the two different logistic regression models. As shown in Figure 2a, the homogeneous logistic regression and heterogeneous logistic regression need about $24\times$ (250s vs 10s) and $17\times$ (177s vs 10s) longer task completion time compared to the simple distributed implementation. Further, we break down the task completion time into CPU time and data transfer time to have a deep view of which factors contributes to the large overhead. The results are shown in Figure 2b and 2c. Both models have a much longer CPU execution time and data transfer time compared to distributed implementation. The results tell us that the state-of-the-art FTL implementation, FATE, still has dramatic computation and data transfer overhead which needs further improvements. In the following sections, we will analyze the reason why computation and data transfer take much longer time in FTL.

### 3.3 CPU Execution Time

For the CPU execution time, it is hard to explicate the reasons for the increase of time simply based on evaluation results. Therefore, we analyze the code and find two possible reasons: 1) more computation workload is introduced due to the data encryption; 2) most of CPU resources are spent on extra tasks such as inter-process communication and memory copies. To verify our assumption, we modify the code to measure the time used in different components of FATE, such as actual model training and extra tasks. Further, we measure the model training time with and without encryption to evaluate the influence of the data encryption.

Figure 3a shows the proportion of the model training time and the extra execution time. Taking heterogeneous logistic regression as an example, only $18\%$ of the CPU execution time (29s out of 160s) is spent on model training. The

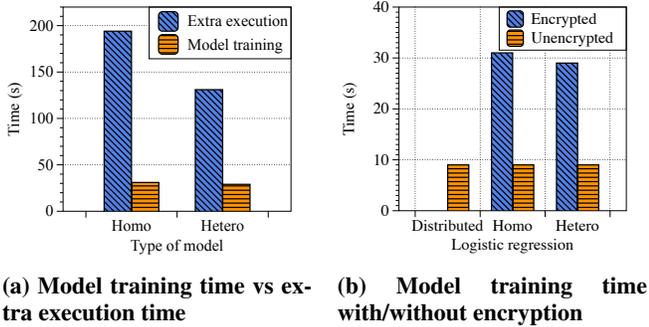

**(a) Model training time vs extra execution time**

**(b) Model training time with/without encryption**

**Figure 3:** Figure (a) shows that only about $18\%$ of the CPU execution time is spent on actual model training. We also explore the influence of data encryption. As illustrated in Figure (b), over $2\times$ more computation workload is caused by current encryption method.

extra CPU consumption causes a considerable performance degradation in the training tasks. To be more detailed, the performance degradation is caused by the following reasons: 1) data exchange and memory copies between different runtimes, such as python and JVM; 2) within the JVM, there are still two processes. One is responsible for creating and publishing the jobs once the data is ready. The other one intensively responses to all kinds of system events and transfers the data to remote parties. The communication between the two processes and the thread scheduling within process affect the efficiency. To conclude, the inter-process communication within machines and memory copies are the major bottlenecks of the current FTL implementation. Fortunately, the bottlenecks are due to none-optimized implementation and can be fixed in the near future.

Our next step is to examine if the data encryption also causes considerable performance degradation. The actual training time with and without data encryption is illustrated in Figure 3b. For both homogeneous and heterogeneous models, the actual training time without encryption is comparable as that in the distributed model. However, when the data is encrypted, the model training time is increased by over 2 $\times$ (29s and 31s vs 9s). Therefore, the requirement for data privacy causes extra computation overhead to FTL because the software-based data encryption needs lots of CPU resources for numeric calculation, which can be a challenge for FTL to be deployed in the production environment. Finally, we are trying to investigate whether advanced hardware can mitigate the above overhead. We adjust the availability of local computation capacity by changing two kinds of resources: 1) Number of virtual CPUs and 2) Available memory. Figure 4 shows the results.

First, we change the number of virtual CPUs from 4 to 10. When more CPUs are available, the CPU time is reduced by up to $12\%$ (from 245s to 215s in the homogeneous model). However, compared to the large performance degradation due to the complexity of FTL, the increased CPU resources are still not enough. The reason why we start from 4 CPUs is that FTL framework has a minimum requirement for the number of CPUs. In our experiments, machines with fewer than 4

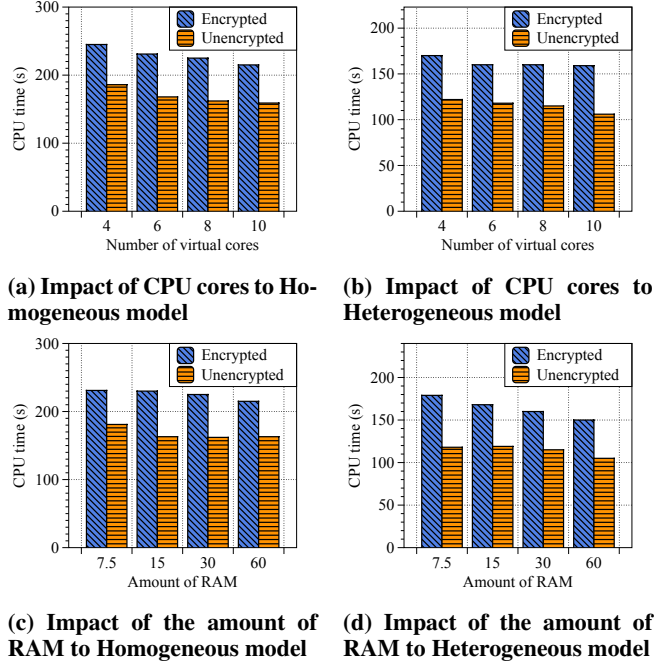

**(a) Impact of CPU cores to Homogeneous model**

**(b) Impact of CPU cores to Heterogeneous model**

**(c) Impact of the amount of RAM to Homogeneous model**

**(d) Impact of the amount of RAM to Heterogeneous model**

**Figure 4:** We adjust the availability of local computation capacity such as CPU cores and the amount of RAM to investigate whether advanced hardware can mitigate the large CPU execution overhead. However, $2.5\times$ CPU resources only reduce the CPU execution time by $12\%$. And $8\times$ amount of RAM only provides $16\%$ improvement.

virtual cores may encounter thread conflicts and fail to access the database that stores the information of other parties such as IP address and receiving ports. Consequently, parties do not know the destination of the data to be transferred, communication among parties breaks off and the whole training process stops.

Similarly, the increase in the amount of available memory does not considerably affect the result. Machines with 60G RAM only provide $16\%$ improvement compared to the ones with 7.5G (from 179s to 150s in the heterogeneous model). The reason is that current implemented models all have small feature spaces and the dataset is also not large enough, so the memory does not play an important role in the overall performance.

### 3.4 Data Transfer Time

In this section, we first explain the reason why data transfer time is also increased by over 16 times in FTL. We have the following two reasons. First, Paillier [Paillier, 1999], a typical algorithm for homomorphic encryption, is used in FATE to achieve the data encryption. Paillier increases the size of a single number to 1024 bits (compared to 64 bits in usual cases). Therefore, after encryption, the overall size of data to be transferred is largely increased. Secondly, FTL needs to modify models to set up privacy protection, which needs more frequent data transfer compared to distributed machine learning. For distributed machine learning, parameter server and worker exchange data at the end of each iteration to update the centralized model. While for FTL, parties need to communicate with each other intensively during even one iteration to encrypt and decrypt data.

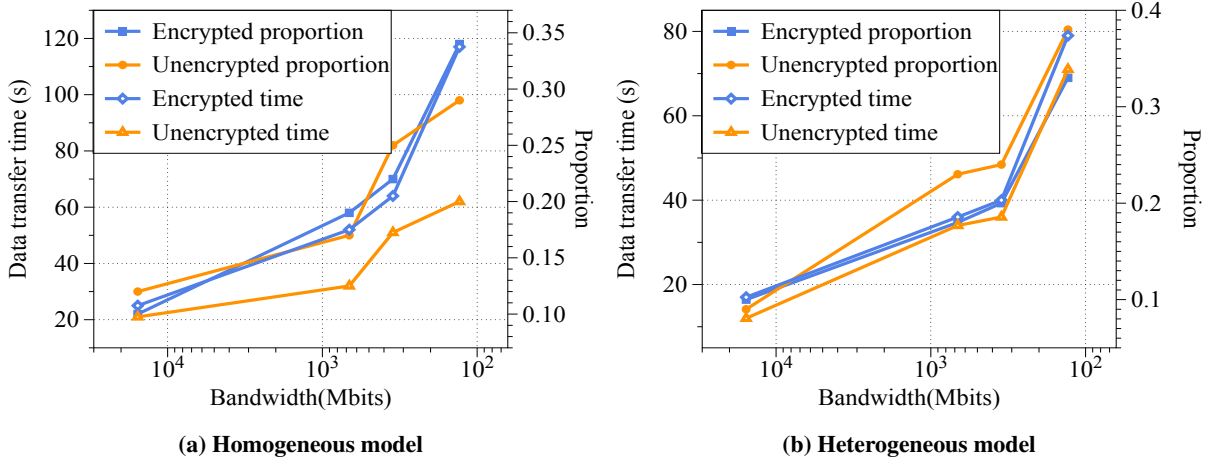

(a) Homogeneous model

(b) Heterogeneous model

Figure 5: We locate parties in different geographical levels to emulate the real scenarios. Different geographical distances among parties cause the variance of bandwidth. Results show that data transfer plays more important roles with bandwidth drops down. When parties are located worldwide, the data transfer time occupies up to 34% of the total completion time.

Moreover, in the production environment, parties may not always be in one data center but may probably have long geographical distances. The bandwidth varies accordingly. To emulate the real cases and evaluate how different geographical levels affects the performance of FTL, we put different parties in different data centers located in the United States, North America and even worldwide. Figure 5 shows the result. When parties are located worldwide, the proportion of data transfer time through the Internet can take up to 34% (117s out of 324s) and 33% (79s out of 239s) of the total completion time in homogeneous and heterogeneous models respectively. To conclude, for FTL to be practical, we need to address the problem that long geographical distance largely degrades the performance of data transfer.

## 4 Bottlenecks and Potential Solutions

From the above evaluations, we summary three bottlenecks and provide possible solutions that can assist us to optimize current FTL frameworks.

1. Inter-process communication (IPC) within machines occupies up to 77% of the total completion time. It is the major bottleneck of the current FTL implementation. Data exchange and memory copies among processes cause extremely high latency. To mitigate this bottleneck, some of the following technologies may help.

   - JVM native memory heap. This can reduce the memory copy between JVM and raw memory.
   - More efficient IPC implementation to reduce the communication overhead. Examples are UNIX domain sockets and JTux.

2. The requirements for privacy protection adds more computation overhead. The problem lies in the unoptimized software-based implementation consumes too much CPU cycles. In recent years, ASIC (such as GPU) or FPGA (netFPGA) are proposed to accelerate machine learning frameworks due to its powerful computation capability. Cloud suppliers continuously support corresponding services [Firestone *et al.*, 2018]. Successful solution has already been proposed by Microsoft to accelerate distributed CNN by implementing deep models in FPGA [Ovtcharov *et al.*, 2015]. These solutions give us the inspirations to consider the possibility of adopting specific hardware for data encryption in FTL infrastructure.

3. Since FTL involves multiple data centers to train a model, network performs a more significant role. The intensive communication among parties generates intensive networking traffic which is easily congested. The requirement for efficient networking urges us to explore the possibilities to deploy advanced congestion control techniques to improve data transfer efficiency. PCC [Dong *et al.*, 2018] may be a viable solution. It can generate more fine-grained and effective congestion control rules and make the network more efficient for long distance data transfer.

## 5 Conclusion

In this paper, we measure the performance of FTL framework based on a real-world application FATE. We compare the performance between FTL and distributed machine learning, which shows a considerable gap. Then we break down the gap into details to explore the factors that influence the performance. Finally, we show three major bottlenecks in current FTL frameworks and propose possible solutions to assist the optimizations in the future.